\documentclass[a4paper,fleqn,usenatbib]{mnras}

\usepackage{newtxtext,newtxmath}
\usepackage{mathptmx}

\usepackage[T1]{fontenc}
\usepackage{ae,aecompl}

\usepackage{graphicx}
\usepackage{txfonts}
\usepackage{epstopdf}
\usepackage{color}

\newcommand{\teff}{$T\rm_{eff}$}
\newcommand{\kms}{km~s$^{-1}$}
\newcommand{\ergcms}{erg cm$^{-2}$\,s$^{-1}$}
\newcommand{\rscvn}{\mbox{RS\,CVn}}
\newcommand{\vsini}{$v \sin i$}
\newcommand{\logg}{\ensuremath{\log g}}
\newcommand{\metal}{\mbox{[A/H]}}

\newcommand{\multi}{\mbox{\it Multi}}

\newcommand{\LTE}{\mbox{LTE}}
\newcommand{\NLTE}{\mbox{NLTE}}

\newcommand{\Halpha}{\mbox{H-$\alpha$}}
\newcommand{\halpha}{\mbox{H-$\alpha$}}
\newcommand{\hbeta}{\mbox{H-$\beta$}}

\newcommand{\cairt}{\ion{Ca}{ii}\,IRT}
\newcommand{\mghk}{\ion{Mg}{ii}\,h\,\&\,k}
\newcommand{\mgii}{\ion{Mg}{ii}}
\newcommand{\caii}{\ion{Ca\,\textsc{II}}}
\newcommand{\nad}{\ion{Na}{i}\,D}
\newcommand{\cahk}{\ion{Ca}{ii}\,H\,\&\,K}

\title[Chromospheric magnetic field from a revised NLTE modeling]{Estimating the chromospheric magnetic field from a revised NLTE modeling: the case of HR\,7428}

\author[I. \,Bus\'a et al.]{I. \,Bus\'a$^{1}$\thanks{E-mail: ebu@oact.inaf.it},
 G. \, Catanzaro$^{1}$,
A. \, Frasca$^{1}$,
M. \, Gangi$^{2,1}$,
M. \, Giarrusso$^{2,1}$,
F. \, Leone$^{2,1}$,\and
M. \, Munari$^{1}$,
C. \, Scalia$^{2,1}$,
S. \, Scuderi$^{1}$
\\
$^{1}$INAF - Catania Astrophysical Observatory,
              Via S. Sofia, 78, 95123 - Catania - Italy\\
$^{2}$Department of Physics and Astronomy, University of Catania, Via S. Sofia, 78, 95123 - Catania - Italy\\
}

\date{Accepted XXX. Received YYY; in original form ZZZ}

\pubyear{2016}

\begin{document}
\label{firstpage}
\pagerange{\pageref{firstpage}--\pageref{lastpage}}
\maketitle


\begin{abstract}

In this work we use the semi-empirical atmospheric modeling method to obtain the chromospheric temperature, pressure, density and magnetic field distribution versus height in the K2 primary component of the \rscvn\ binary system HR\,7428.
While temperature, pressure, density are the standard output of the semi-empirical modeling technique, the chromospheric magnetic field estimation versus height comes from considering the possibility of not imposing hydrostatic equilibrium in the semi-empirical computation. The stability of the best non-hydrostatic equilibrium model, implies the presence of an additive (toward the center of the star) pressure, that decrease in strength from the base of the chromosphere toward the outer layers. Interpreting the additive pressure as magnetic pressure we estimated a magnetic field intensity of about 500~gauss at the base of the chromosphere.
\end{abstract}

 \begin{keywords}
 modeling atmosphere, stellar activity, \NLTE\ radiative transfer, chromosphere, magnetic field
\end{keywords}

\section{Introduction}
HR\,7428 (=~V~1817~Cygni) is a bright (V=6.3) long-period (108.578d) spectroscopic \rscvn\ binary composed by a K2 II-III star and a main sequence A2 star \citep{Parsons&Ake87}. The magnetic activity of the system is well known: \cahk\ emission was first reported by \cite{Gratton50}, by a detailed analysis of photometric observations \cite{Hall_etal90} were able to detect starspot signatures on the K2 primary star. It is now well established that the stellar atmosphere of cool stars is characterized by that temperature gradient inversion. An inversion explained in the framework of the magnetic activity theories, but not yet definitively understood.
\begin{figure*}
\begin{flushleft}
\includegraphics*[width=1.0\textwidth]{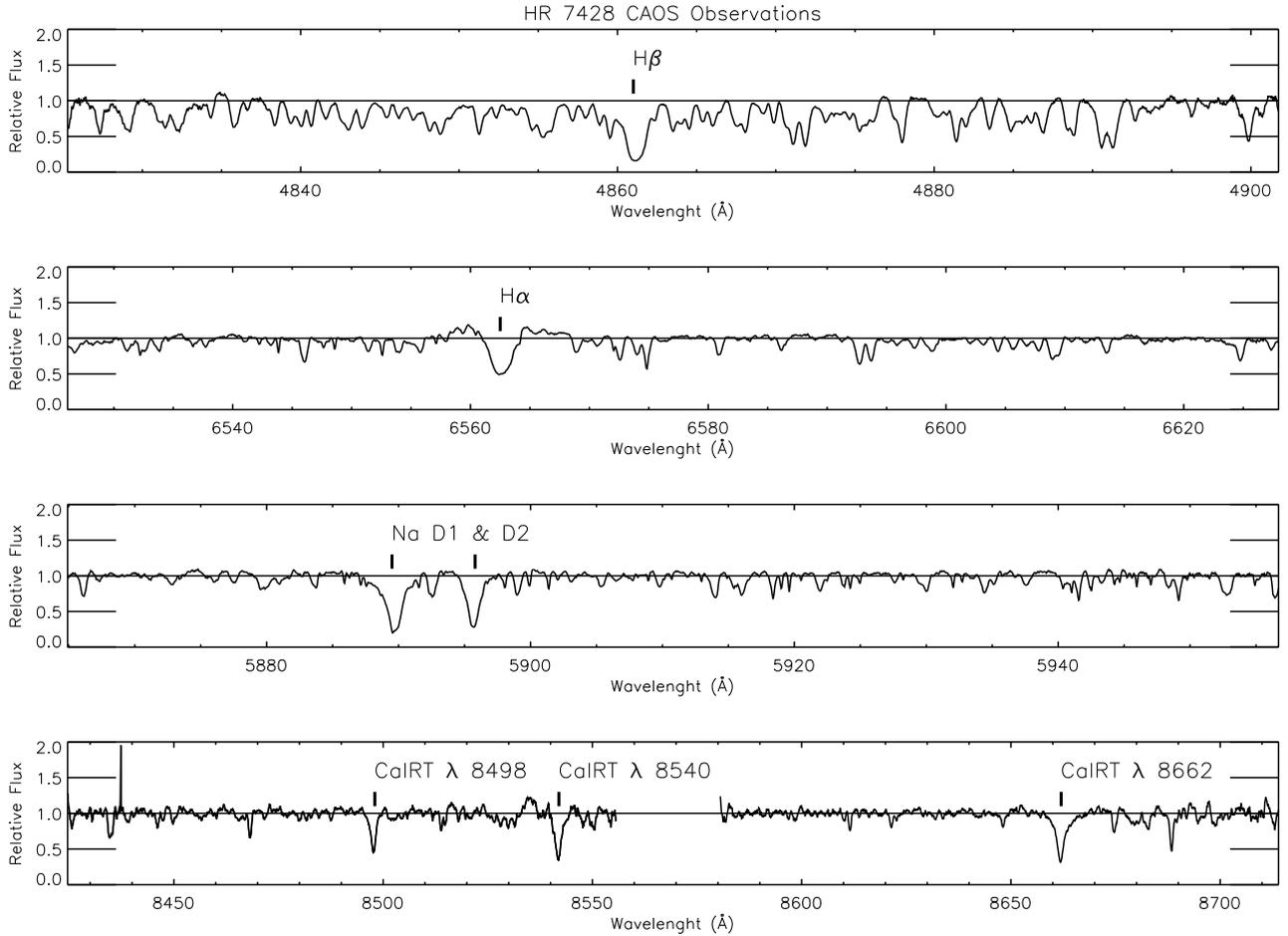}
\end{flushleft}
\caption{HR7428 normalized spectrum obtained using the new Catania Astrophysical  Observatory  Spectropolarimeter (CAOS) showing the line profiles of \hbeta, \nad, \halpha\ and \cairt\ lines.}
\label{caos}
\end{figure*}
Late-type stars, with \halpha\ in emission, show also a fairly stable chromospheric emission outside flares, see. e.g., \cite{Byrne_etal98}. This reinforces the hypothesis that chromospheres are globally in a quasi-stationary state, modulated mainly by the stellar activity cycle, whose temperature-density structure results from the balance between global dissipation of
non-radiative energy and radiative cooling \citep{Kalkofen_etal99}.

Most of what we know about stars and systems of stars is derived from an analysis of their radiation and this knowledge will be secure  only as long as the analytical technique is physically reliable. A well tested technique to get information on physical properties of chromospheric layers of active stars is the NLTE radiative transfer semi-empirical modeling: for different temperature vs height distributions, the NLTE populations for hydrogen are computed, by solving simultaneously the equations of hydrostatic equilibrium, radiative transfer, and statistical equilibrium. The emerging profiles for some chromospheric lines and continua are computed and compared to the observations. Then, the modeling is iterated until a satisfactory match is found. (see, e.g.,\cite{Vernazza_etal81}; \cite{Fontenla_etal93}). These models are built to match the observations in different spectral features, and make no assumption about the physical processes responsible for the heating of the chromosphere, but can be used as constraints for these processes. The obtained models describe the variations of the essential physical parameters, in particular the temperature, pressure and electron density across the outer atmosphere, and give information on its ''mean'' state, both temporally and spatially. 

The most important problem of this approach lies in the uniqueness of the solution.  In fact, knowing that a particular atmosphere
would emit a line profile like the one we observe for a given star does not imply that the star has indeed this atmospheric structure, since we do not know
whether some other atmosphere would produce the same profile.  To solve, or
at least to reduce, this problem, the modeling has to be based on several spectral features, with different regions of formation. The  amount and the kind of diagnostics used  to  build  an atmospheric  model  is in fact, very  important, by  combining several spectral lines that are formed at different but overlapping depths in  the  atmosphere, we  can  obtain  a  more  reliable  model \citep{Mauas_etal06}. 

Certainly the best known semiempirical model is the one for the average Quiet-Sun, Model C by \cite{Vernazza_etal81}. 
Semiempirical modeling was successfully applied also to the atmospheres of cool stars. An extensive modeling of dM stars, has been done, starting  with the work by \cite{Cram&Mullan79}, \cite{Short&Doyle98}, \cite{Mauas&Falchi94}, and \cite{Mauas_etal97}. Furthermore, a rich history of semiempirical chromospheric modeling has been also carried out for cool giant and supergiant stars (see, e.g.,\cite{Kelch_etal78}; \cite{Basri_etal81}; \cite{Luttermoser_etal94}). In cool stars the application of NLTE semi-empirical chromospheric modelling can be based on optical and ultraviolet (UV) observations. This is because lines such as the  \halpha\ \nad\ \cairt\ become dominated by electron-collision excitation processes, which make them effective chromospheric diagnostics \citep{Houdebine_96}. The possibility of using \Halpha\ profile as a diagnostic of stellar chromospheres was discussed in detail by \cite{Cram&Mullan79} and \cite{Mullan&Cram82} in terms of control of the source function by photons or collisional processes. \Halpha\ is observed in active stars in a wide variety of shapes and sizes; when the effective temperature is low enough, the collisional control of the \Halpha\ source function becomes possible over a wide range of chromospheric pressures, and, under this conditions, \Halpha\ can be a good chromospheric diagnostic. \mghk\ UV lines, due to their large opacity, provide excellent diagnostic over a wide range of heights of the outer chromospheric layers \citep{Uitenbroek92}, and  \cairt\ triplet is a constraint for the shape of the middle chromosphere from the temperature minimum up to the plateau \citep{Andretta_etal2005}.

Here we applied the NLTE semi-empirical chromospheric modelling to the K2 star of HR~7428 binary system basing the analysis on the \halpha, \hbeta, \nad, \cairt\ triplet, \mghk\ lines and UV continuum diagnostics.
\label{Sec:intro}

\section{Data acquisition and reduction}
\label{data}
\begin{figure}
\begin{flushleft}
\includegraphics*[width=0.5\textwidth]{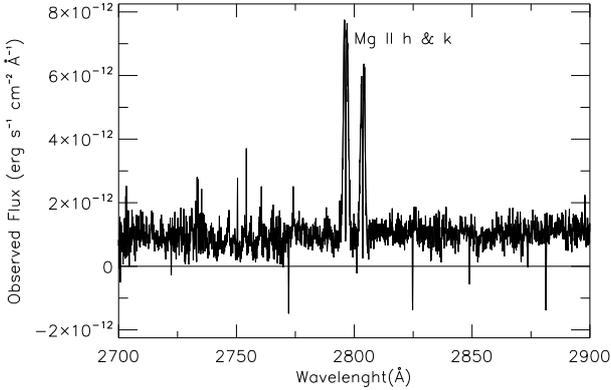}
\caption{HR7428 high-resolution flux calibrated spectrum in the wavelength of the \mgii\ line profile obtained by International Ultraviolet Explorer (IUE).}
\label{mgii_iue}
\end{flushleft}
\end{figure}
\begin{figure}
\begin{flushleft}
\includegraphics*[width=0.5\textwidth]{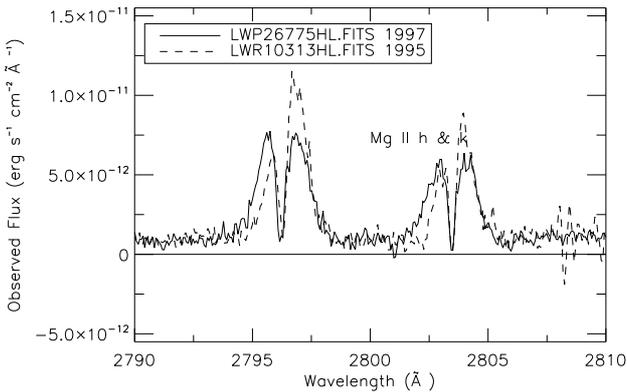}
\caption{Comparison of two IUE spectra of the system, obtained in different dates: LWP26775HL.FITS obtained in November 1997 and LWR10313HL.FITS obtained in December 1995.}
\label{mgvar}
\end{flushleft}
\end{figure}
 \halpha\ \nad\  \hbeta\ \cairt\  spectroscopic observations of HR\,7428 were carried out at the 91-cm telescope of Catania Astrophysical Observatory, ''M. G. Fracastoro'' station (Serra La Nave, Mt. Etna, Italy), using the new Catania Astrophysical  Observatory  Spectropolarimeter (CAOS) which is a fiber fed, high-resolution, cross-dispersed echelle spectrograph (\cite{leone_etal16}; \cite{Spano_etal04}, \cite{Spano_etal06})

The spectra were obtained in September 2015. Exposure times have been tuned in order to obtain a signal-to-noise ratio of at least 200 in the continuum in the 390-900 nm spectral range, with a resolution of R=$\frac {\lambda}{\Delta(\lambda)}$~=~45,000, as measured from ThAr and telluric lines.

Echelle IRAF packages have been used for data reduction, following
the standard steps: bias subtraction, background subtraction, trimming, flat-fielding and scattered light subtraction, extraction for the orders, and
wavelength calibration. Several thorium lamp exposures were obtained
during each night and then used to provide a wavelength calibration of the observations.
Each spectral order was normalized by a polynomial
fit to the local continuum. 
A reduced spectrum, in the wavelength ranges of the line of interest is shown in
  Fig.~\ref{caos}. In the plotted spectrum the average S/N obtained at the continuum
close to \halpha\ is about 200. 

\mghk\ spectroscopic observations have been obtained, in 1997, by the IUE satellite. 
IUE spectra have been corrected for interstellar extinction. 
According to the Hipparcos distance d~=~323$\pm$52~pc and to the typical
value of 1 Mag per kilo-parsec for interstellar extinction we adopted for the
computation A(V)~=~0.32. Assuming the standard reddening law $A(V)$~=~3.1 $\times$ E(B-V), a color excess E(B-V)~=~0.10 has been derived. IUE spectra have been de-reddened according to the selective extinction function of \cite{Cardelli_etal89}. 

The spectral resolution is about 0.2 \AA\ for the \mgii\ region.
The IUE flux calibrated HR7428 spectrum, is shown in  Fig.~\ref{mgii_iue}. 

Unfortunately, we have no simultaneous UV and optical observations. 
Therefore we have to take into account the activity-variability that could affect the HR7428 system,  due to different levels of the stellar activity and/or different distribution of the active regions on the visible surface, in different times. In order to have an estimate of how much UV data are affected by activity-variability  we have compared two IUE spectra of the system, obtained in different dates: LWP26775HL.FITS obtained in November 1997 and LWR10313HL.FITS obtained in December 1995 (see Fig.~\ref{mgvar} where the two \mghk\ IUE spectra are shown with a shift in wavelength in order to overlap.)
$1.060 \times 10^{-12} \pm 1 \times 10^{-15}$
We find that we can neglect the long term variability as far as the  UV continuum is concerned, the two spectra show in fact the same continuum mean value (\mgii\ Continuum Flux(1995)~=~$1.06 \times 10^{-12} \pm 1 \times 10^{-14}$ \ergcms, \mgii\ Continuum Flux(1997)~=~$1.060 \times 10^{-12} \pm 7 \times 10^{-15}$  \ergcms). As far as the \mghk\ line profiles are concerned, we measure approximately equivalent observed fluxes at Earth ($Flux_{\small \mghk}$(1995)~=~$3.33 \times 10^{-11} \pm 3 \times 10^{-13}$ \ergcms, $Flux_{\small \mghk}$(1997)~=~$3.20 \times 10^{-11} \pm 3 \times 10^{-13}$ \ergcms), but the profile shape is quite different as shown in Fig.~\ref{mgvar}, most probably indicating a different distribution of active regions in different times, that we cannot take into account. Therefore an higher weight will be done to the best fit of the UV continuum with respect to the \mghk\ line profile.

\section{Computational method}

Historically, K2 giants were classified as "non-coronal" stars, however, in the late 1990s, evidence for transition region emission was detected for Arcturus (K2 III) and Aldebaran (K5 III) \citep{Ayres_etal03} and for other representative ''non-coronal'' red giants (see, e.g.,\cite{Ayres_etal97}; \cite{Robinson_etal98}) Therefore, here, the  atmospheric model of the K2 primary magnetic active component has been built computing a photospheric model, a chromospheric model and a transition region model and joining the three together; we assume a plane-parallel geometry in our modeling efforts.

\subsection{Photospheric Model}
Taking into account the \cite{Marino_etal01} physical parameters (see Table~\ref{mar_par}) we selected from the Castelli LTE synthetic spectra database (http://www.oact.inaf.it/castelli/castelli/grids.html), a spectrum with parameters \logg\~=~2.0, 
\teff\~=~4400~K and solar metalicity that describes the photospheric contribution of  the K2 primary component, and a spectrum with parameters  \logg\~=~4.0, \teff\~=~9000~K and solar metalicity that describes the flux contribution of the A2 secondary component of the binary system HR~7428.
\begin{table}
\caption{HR\,7428 K2II-III and A2 components \citep{Marino_etal01}.}
\label{mar_par}
\begin{center}
\begin{tabular}{ccc}
\hline
\\
Element & Primary (cooler K2II-III) & Secondary (hotter A2) \\
\hline
\\
R  &$ 40.0 \pm\ 6.5 R_{\odot} $ & $ 2.25 \pm\ 0.5 R_{\odot} $ \\
\teff\ &$ 4400~K \pm\ 150~K $&$ 9000~K \pm\ 200~K $\\
\logg\  &$ 2.0 \pm\ 0.5 $ & $ 4.0 \pm\ 0.5 $ \\
\\
\hline
\end{tabular}
\end{center}
\end{table}
\begin{figure}
\begin{flushleft}
\includegraphics*[width=0.5\textwidth]{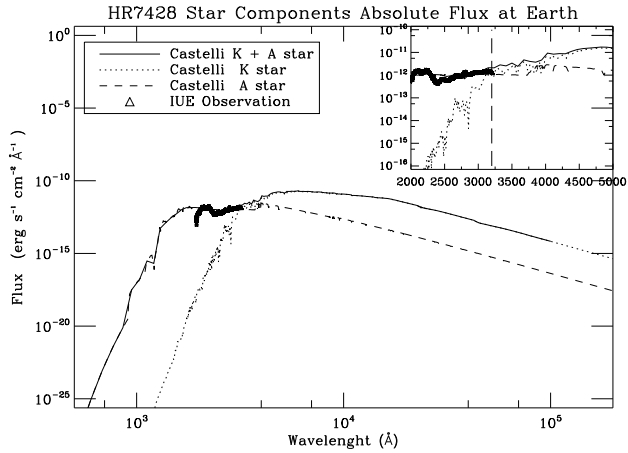}
\end{flushleft}
\caption{HR7428 Castelli \LTE\ synthetic photospheric fluxes at Earth for the K star (dotted line) and A star (dashed line) where the conversion factor $R^2$/$d^2$ has been calculated for the K and A stars according to the parameters given in Tab.\ref{mar_par}. The plot zoom around 2000-5000~$ \AA\ $ shows that before $ \lambda\ $~=~3200~$ \AA\ $ the A star dominates the continuum emission, while for wavelength greater than $ \lambda\ $~=~3200~$ \AA\ $ we can neglect the A star contribution, the continuum is in fact dominated by the K star.} 
\label{pesostellaa}
\end{figure}
In Fig.~\ref{pesostellaa} the fluxes at Earth of the two LTE models are shown together with their sum. The original fluxes have been converted to flux at Earth by the conversion factor $R^2$/$d^2$ that has been calculated for the K and A star according the parameters given in Tab.\ref{mar_par}. From the plot we conclude that, for \NLTE\ radiative transfer calculations of the \hbeta, \nad, \halpha, \cairt\ triplet profiles of the binary system, we can neglect the A2 star contribution. In fact, from Fig.~\ref{pesostellaa} we can see that for wavelength longer than 3200 \AA, the continuum flux of the binary system is dominated by the K star. This is in agreement with \cite{Marino_etal01} that find the A2 star contribution to the total flux in the \halpha\ region is only 4\%. On the other hand, in order to calculate the \NLTE\ \mghk\ (wavelength lower than 3200 \AA) HR\,7428 line profile, we have to take into account the A2 star contribution. In the \mghk\ spectral range in fact (triangles show the observed IUE \mghk), the observed continuum is dominated by the A star contribution.

Therefore, we computed the A2 secondary component atmospheric model
by the ATLAS9 code \citep{Kurucz93} using the parameters \logg=4.0, \teff= 9000~K, \metal=0.0 .

The atmospheric model of the K2 primary magnetic active component has been built computing separately a photospheric model, a chromospheric model and a transition region model and joining the three together.

The K2 photospheric model was computed using the ATLAS9 code and the parameters \logg=2.0, \teff= 4400~K, \metal=0.0  for the K2 primary component.

\subsection{Transition Region Model}

A first estimation of a plane-parallel model for the lower transition region of the HR\,7428 K2 primary component, has been built using the method of the Volumetric Emission Measure. The use of Emission Measure techniques to construct transition region models is well established (see for example \cite{JordanBrown81}, \cite{Harper92}), the flux at the star, for lines forming at temperature $T_e$~$\approx$~$10^5$~K is in fact dominated by collisions. This results in emission lines that are optically thin and with a contribution function sharply picked in temperature, that is, typically formed over a temperature range of $\Delta\log(T_e)$~=~0.30 . 

The above conditions allow to determine the temperature gradient as a function of the averaged Emission Measure over $\Delta\log(T_e)$~=~0.30  that we indicate as $EM_{0.3}$.

By imposing hydrostatic equilibrium, including turbulent pressure, the transition region model can be obtained combining the temperature gradient as a function of $EM_{0.3}$ and the pressure variation from the equation of hydrostatic equilibrium (see,e.g., \citet{Harper92}). 

The expression for the temperature gradient combined with the equation of hydrostatic equilibrium gives the relationship:

\begin{equation}
\small
\begin{tabular}{l}
$P_{Tot}^2$($T_2$) - $P_{Tot}^2$($T_1$)~=~2 $\times$  $(1.4)^2 \, m_p \,g \,k  \,\times$  \\
$\times \, \int_{T_1}^{T_2}{[EM_{0.3} \times \left(1 \dotplus  1.1 x \right) + \frac{1.4\,x\,  m_p \, v_{turb}^2  EM_{0.3}}{2\, k\, T} ] dT}$\\
\end {tabular}
\label{eqintegr}
\end {equation}
\begin{figure}
\begin{flushleft}
\includegraphics*[width=0.5\textwidth]{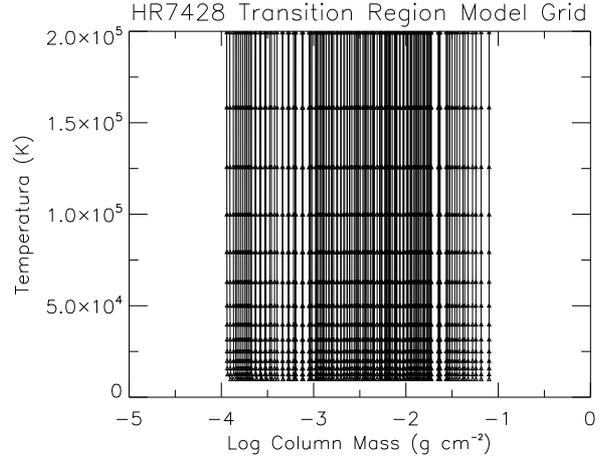}
\end{flushleft}
\caption{Grid of transition region models used for the study of HR7428 atmosphere. The grid has been built as described in the text.}
\label{grigliatr}
\end{figure}
\noindent 
where $m_p$ is the $H^{+}$ mass, $k$ is the Boltzmann constant, $v_{turb}$ is the turbulent velocity and x$\equiv$$N_H/N_e$.

The Eq. \ref{eqintegr} together with an estimate of electron density and turbulent velocity in a layer, allows to find the pressure as a function of temperature for the TR model. The total particle density ($N_{tot}$), gas pressure
($P_G$), turbulence pressure ($P_{turb}$) and electron density ($N_e
$) are then obtained according to the following relations
\begin{equation}
\begin{tabular}{l}
\label{tante}
$N_{tot}$~=~$ \sqrt{P_{Tot}}$ \,(k\, T + 0.5 $m_p$ \,$v_{turb}^2$)\\
$P_G$~=~$ N_{tot}$\,k\,T \\
$P_{turb}$~=~$ 0.5 N_{tot}$ \,$m_p$ \,$\mu$ \,$v_{turb}^2$\\
$N_e$~=~$ N_{tot}$ / (1.+1.1\,x) \\ 
\end {tabular}
\end {equation}
where $\mu \equiv  \frac{1.4 N_H/N_e}{\left(1 \dotplus 1.1 N_H/N_e \right)}$ is the molecular weight.

We used the equations above in order to build plane-parallel models for the lower transition region of the HR\,7428 K2 primary component. 

In order to have an estimate of Emission Measures for the HR\,7428 binary system, we used, the
values of Volumetric Emission Measure ($VEM$) vs \teff\ measured by \cite{GriffithsJordan98} for another \rscvn\ system, HR\,1099 primary component ($R_{HR1099 \,K1IV}$~=~3.9 $R{_\odot}$),  opportunely scaling them, in order to take into account the bigger radius ($R_{K2}$~=~40 $R{_\odot}$) of HR\,7428 K2 star, according the formula $EM_{0.3}$~=~$VEM/4\pi R^2$ \citep{Brownetal91} and the parameters of Table, \ref{mar_par}.

From the estimated $EM_{0.3}$ values, a grid  of 160 transition region models, shown in Fig.~\ref{grigliatr} has been built by means of equations \ref{eqintegr} and \ref{tante} using a grid of 23 values of electron density at the fixed temperature $T_e$~=~50000~K obtained scaling of a factor from 0.5 up to 30 the values of electron density at 50,000~K measured in HR\,1099 ($N_e$~=~5~$\times 10^{+11}$cm$^{-3}$). For each of these 23 electron density values, a grid of seven values of the turbulent velocity in the layer with $T_0$~=~$10^4$~K, $v_{turb}\left(T_0 \equiv ~10^4~ \mbox{K} \right)$~=~10, 20, 30, 40, 50, 60, 70  \kms\ has been considered for the calculation of turbulent velocity distribution according the empirical law by \cite{GriffithsJordan98}  $v_{turb}$(T)=$v_{turb}\left(T_0\right) \times \left(T/T_0\right)^{1/4}$ between $\log{(T)}$~=~4.0 and $\log{(T)}$~=~5.3

These transition region models provide the upper boundaries for the radiative transfer calculations of the chromospheric models, while the adopted photospheric model provides the lower boundaries of the chromospheric models.

\subsection{Chromospheric Model}
\begin{figure}
\begin{flushleft}
\includegraphics*[width=0.5\textwidth]{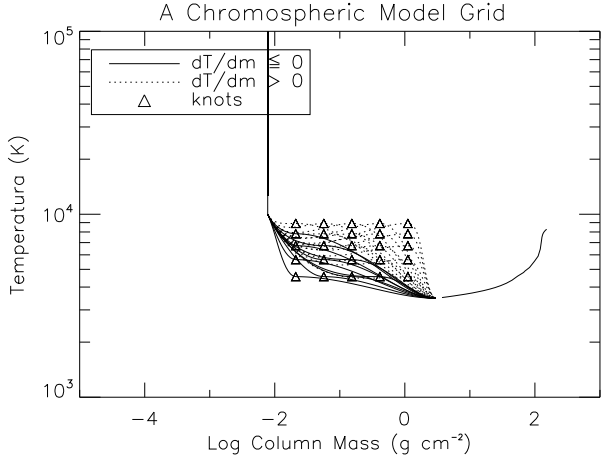}
\end{flushleft}
\caption{Grid of chromospheric structures obtained by spline interpolation between a fixed Transition Region model and a fixed Photospheric model, using as free parameters a grid of 25 knots point. The figure shows in dotted lines the models that do not satisfy the negative gradient condition $dT/dm \leq 0$.}
\label{25modello_chr}
\end{figure}
The term 'chromosphere' indicates the region above a stellar photosphere where non-radiative heating processes (either magnetic or acoustic) become important in the energy balance and where hydrogen is partially ionized. Since hydrogen is nearly completely ionized when temperatures reach 20,000-30,000~K, we consider this the top of chromosphere. Once hydrogen is nearly completely ionized, mechanical heating produces a steep thermal gradient because there is no cooling channel as effective as hydrogen, and \caii, \mgii\ have disappeared before hydrogen is ionized \citep{linsky_01}.

Chromospheres lies in the difficult regimes of non-LTE and non-ionization equilibrium. While photosphere can be calculated in detail when having \logg, \metal\ and \teff, and transition region can be constrained by observations because lines form in optically thin conditions as seen before, for the chromospheric layers we have no constraints, lines are optically thick, and the chromospheric model has to be based upon spectral diagnostic methods by means of semi-empirical modeling technique, that is changing an hypothetic model iteratively, in order to match as many as possible observations that form in the chromospheric layers. We can only suppose that the temperature increases with a smooth thermal gradient because, the non-radiative sources heats the plasma producing bright emission in the \halpha, \cairt, \mghk\ which together are the dominant cooling channels of the chromosphere. In the Sun the chromosphere goes from $\approx$~600~km and $\approx$~2000~km above the photosphere and is characterized by temperature gradient sign change, where temperature smoothly grow from $\approx$~4500~K up to $\approx$~15000~K over which the gas density changes by several orders of magnitude due to a sharp decrease of turbulent pressure  while the electron density only smoothly decrease. 

The solar chromosphere has been described for the first time by \citet{Vernazza_etal73} by means of a one-component model of the solar atmosphere, including in that model the photosphere, chromosphere and transition zone, and then in detail, by \citet{Fontenla_etal99}, who using the PANDORA NLTE radiative transfer code determine semiempirical models for seven semiempirical models for sunspots, plages, network, and quiet atmosphere constructed to reproduce observed emergent intensities and profiles at wavelengths from the UV to radio wavelengths.


While \citet{Vernazza_etal73} determine the chromospheric model of the Sun by adjusting the temperature as a function of height to that distribution that gives best agreement between synthesized and observed line spectra, here we have built a wide grid of chromospheric models from which to look for the best model by means of a $\chi^2$ minimization selection method.

In particular, for each one of the 160 transition region model and for each of nine values of  $T_{min}$ in the range between $\sim$2800~K and 4200~K that we have chosen with a step of less than 200~K as points where to cut the photospheric Kurucz model, a grid of 25 chromospheric models are generated by a smooth spline interpolation  between the photosphere and the transition region using as free parameters  a grid of $5 \times 5$ interpolation knots, (see Fig.~\ref{25modello_chr}, where the grid of 25 chromospheric models is shown for a fixed TR model and a fixed minimum of photospheric model $T_{min}$. We impose the chromospheric structures to have a monotonic temperature dependence on Column Mass ($dT/dm \leq 0$). Fig.~\ref{25modello_chr} shows how the temperature gradient constraint, strongly cuts the number of useful model of the grid, for example in the 25 model grid of the figure, only eight models satisfy the gradient constraint and can be used in the analysis.

The total grid of models includes 160*9*25=36225 models, and only 15691 satisfy the $dT/dm \leq 0$ constraint and have been considered in the \NLTE\ radiative transfer.

\subsection{Computation: applying or not hydrostatic equilibrium equations}
\begin{figure*}
\includegraphics*[width=0.9\textwidth]{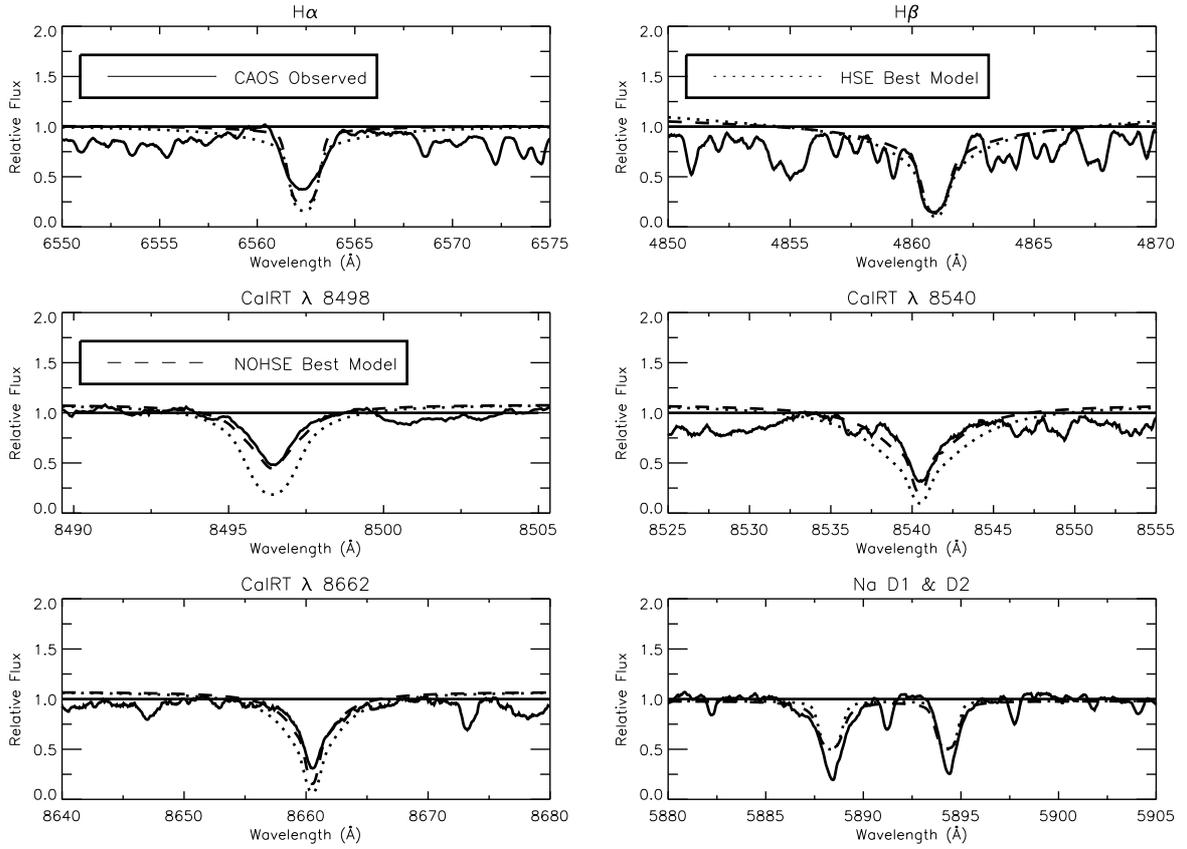}
\caption{\NLTE\ \halpha, \hbeta, \cairt, \nad\ normalized profiles computed for the best $HSE$ (dotted line) and the best $NOHSE$ (dashed line) models compared with observations.}
\label{righe}
\end{figure*}
The coupled equations of radiative transfer and statistical equilibrium were
solved using the version 2.2 of the code \multi\ \citep{Carlsson86}, for the  \ion{H}{}, \ion{Ca}{}, \ion{Na}{}, \ion{Mg}{} atomic models.
The \ion{H}{} atomic model incorporates 16 states of \ion{H}{}, with 84 {\it b-b} 
and 9  {\it b-f} transitions. The \ion{Ca}{} atomic model incorporates 8 states 
of \ion{Ca}{i}, the lowest 5 states of \ion{Ca}{ii} and the ground state of 
\ion{Ca}{iii}, 9 {\it b-b} and 13 {\it b-f} transitions are treated in detail. 
The \mgii\ atomic 
model is made of 3 states \ion{Mg}{i}, the lowest 6 states of \mgii and the ground state of \ion{Mg}{iii}, 9 {\it b-b} and 9 {\it b-f} transitions are treated in detail. 
The \ion{Na}{} atomic model incorporates 12 levels: 11 levels of \ion{Na}{i} and the ground state of \ion{Na}{ii} and 29 {\it b-b} and 11 {\it b-f} transitions are treated in detail. 
The opacity package included in the code takes into account free-free opacity, Rayleigh scattering, and bound-free transitions from hydrogen and metals, We included the line blanketing contribution to the opacity using the method described in \cite{Busa_etal01}. 
\begin{figure}
\includegraphics*[width=0.5\textwidth]{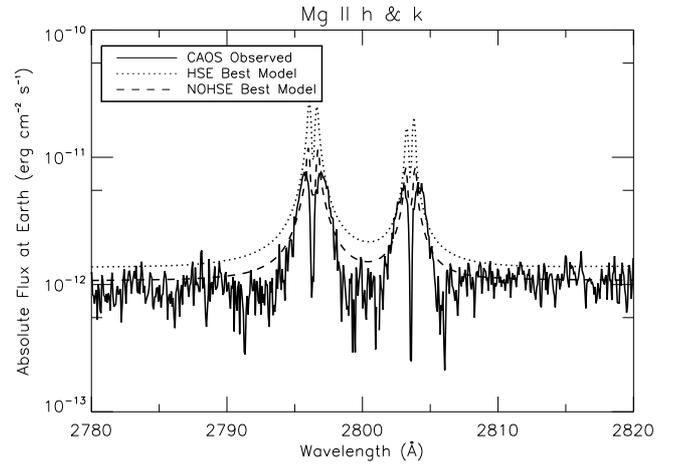}
\caption{\mghk\ absolute flux profiles computed from the best $HSE$ and the best $NOHSE$ atmospheric models compared with the IUE observation.}
\label{mg}
\end{figure}

As a first step we imposed hydrostatic equilibrium to the hydrogen, that is, we calculated the hydrogen radiative transfer  and the equation of hydrostatic equilibrium consistently. 
In detail, starting with the LTE hydrogen populations for the electron pressure and density calculation, hydrogen is iterated to convergence. Then hydrostatic equilibrium $dP_{Tot}$~=~$\rho g\, dh$ is solved and electron pressure and hydrogen populations are updated, the loop continues until we obtain a convergence.
The electron density obtained  is then used to solve the \ion{Ca}{}, \ion{Mg}{}, and \ion{Na}{} radiative transfer and statistical equilibrium equations. The population densities obtained from the \ion{H}{} calculation are used to obtain the background NLTE source function in the \ion{Ca}{}, \ion{Mg}{}, and \ion{Na}{} calculations. This computation modifies the initial grid because a new column of electron density is obtained, we find that only 2052 models converge to a solution  and we call them HSE models.
\begin{figure*}
\includegraphics*[width=0.9\textwidth]{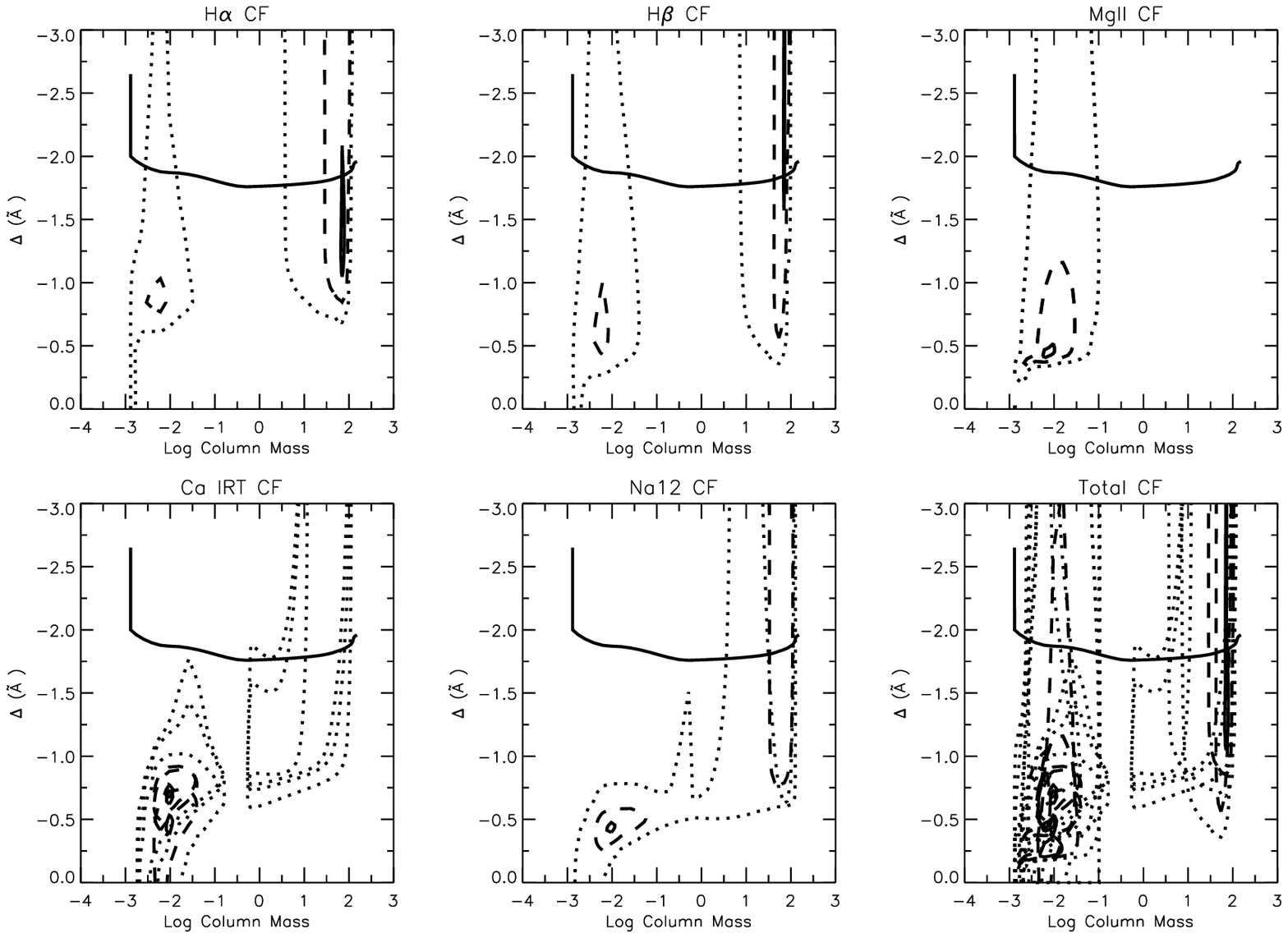}
\caption{Contribution functions of the \halpha, \hbeta, \mghk, \cairt, \nad\ lines for the best $NOHSE$ model. The maximum of the contribution function is set equal to one. The contour plot indicates
the fractions 0.9 (solid), 0.3 (dashed) 0.01 (dotted).  The atmospheric model is plotted as a solid line in each $CF$ plot. The last plot, where all lines contribution functions are plotted  together, indicates that our selected diagnostics mainly constraint the whole atmosphere from the photosphere (\halpha\ and \hbeta\ wings), and chromosphere (all the lines) up to the transition region (\mghk\ cores) giving strength to our semiempiral modeling.}
\label{cntrbfnohse}
\end{figure*}

We also considered not imposing  hydrostatic equilibrium that means fixing the electron densities to the ones of the original grid.
In this case we consider all the 15691 models that only satisfy the $dT/dm \leq 0$ constraint and we call these models, $NOHSE$ models. 

Even if an atmospheric model in non-hydrostatic equilibrium is not realistic and should be rejected, here we assume the $NOHSE$ models as good as the $HSE$ ones. Of course hydrostatic equilibrium has to happen in the star, but, because we are not taking into account all the pressure contributions to the total pressure, therefore we can assure that also the $HSE$ models should be rejected, or, at least can be, good or not, just like the $NOHSE$ models. In a sense, we can assume that, imposing $HSE$, the resulting model is not in hydrostatic equilibrium if a contribution to the pressure is missing in the $dP_{Tot}$~=~$\rho g\, dh$ equation. 
\begin{figure*}
\includegraphics*[width=0.9\textwidth]{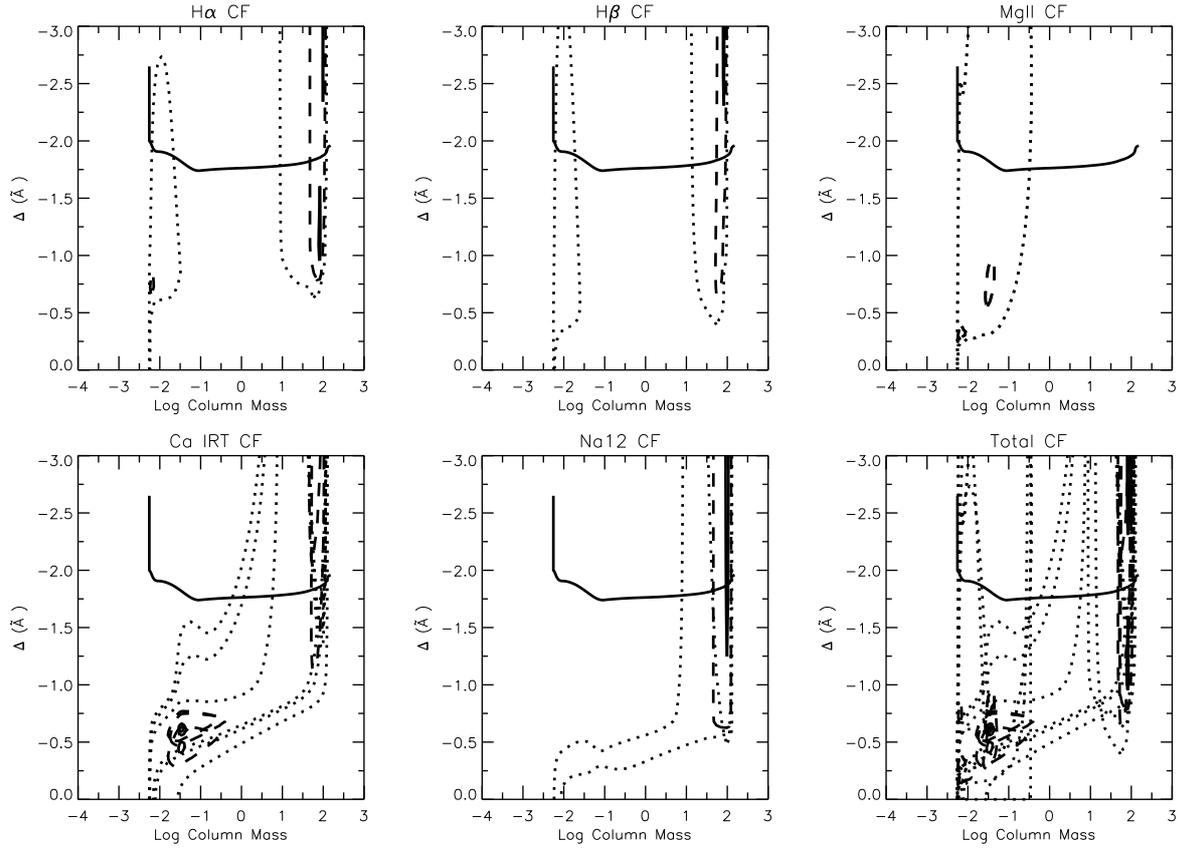}
\caption{Contribution functions of the \halpha, \hbeta, \mghk, \cairt, \nad\ lines for the best $HSE$ model. Like for the $NOHSE$ model (see Fig.~\ref{cntrbfnohse} also for the $HSE$ model the outer atmosphere is enough constrained by our diagnostics.)}
\label{cntrbfhse}
\end{figure*}
\begin{figure}
\includegraphics*[width=0.5\textwidth]{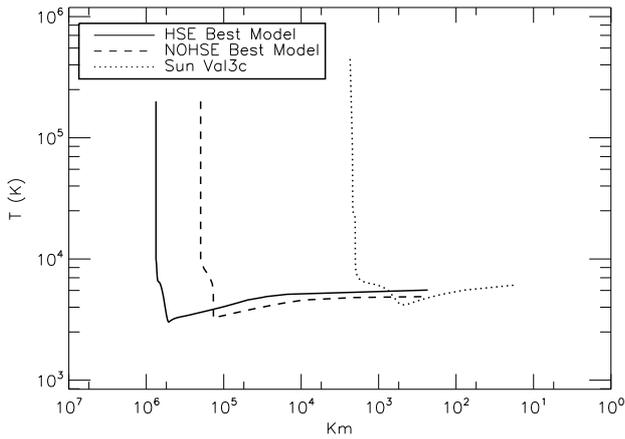}
\caption{Best atmospheric $HSE$ and $NOHSE$ models for the mean K2 primary component of the \rscvn\ system HR~7428 compared with the Solar Val~3c model.The plot describes the distribution of temperature versus height. Height is given in kilometers measured above a zero point where $\tau_{5000}$~=~1.}
\label{bestmodel}
\end{figure}
And this is the case of cool active stars.
We know in fact that the atmospheres of active stars are permeated by magnetic fields that emerge from deeper layers. With increasing height we might expect the structure to be more greatly influenced by magnetic fields, since the energy density of the magnetic fields should fall off more slowly than the energy density of the gas (this is the case of solar atmosphere where $\beta$~=~$8 \pi N K T_e / |(B)|^2$ is $\ge 1$ in photosphere and $<< 1$ in  transition region layers). Therefore, magnetic pressure should be added in the computation of the total pressure and is not. This lack together with the lack of any other possible contribution not yet identified, let us to say that the electron densities and the hydrogen populations obtained from imposing $HSE$ differ from the ones we would have obtained introducing a magnetic filed contribution. 
\begin{figure*}
\includegraphics*[width=0.9\textwidth]{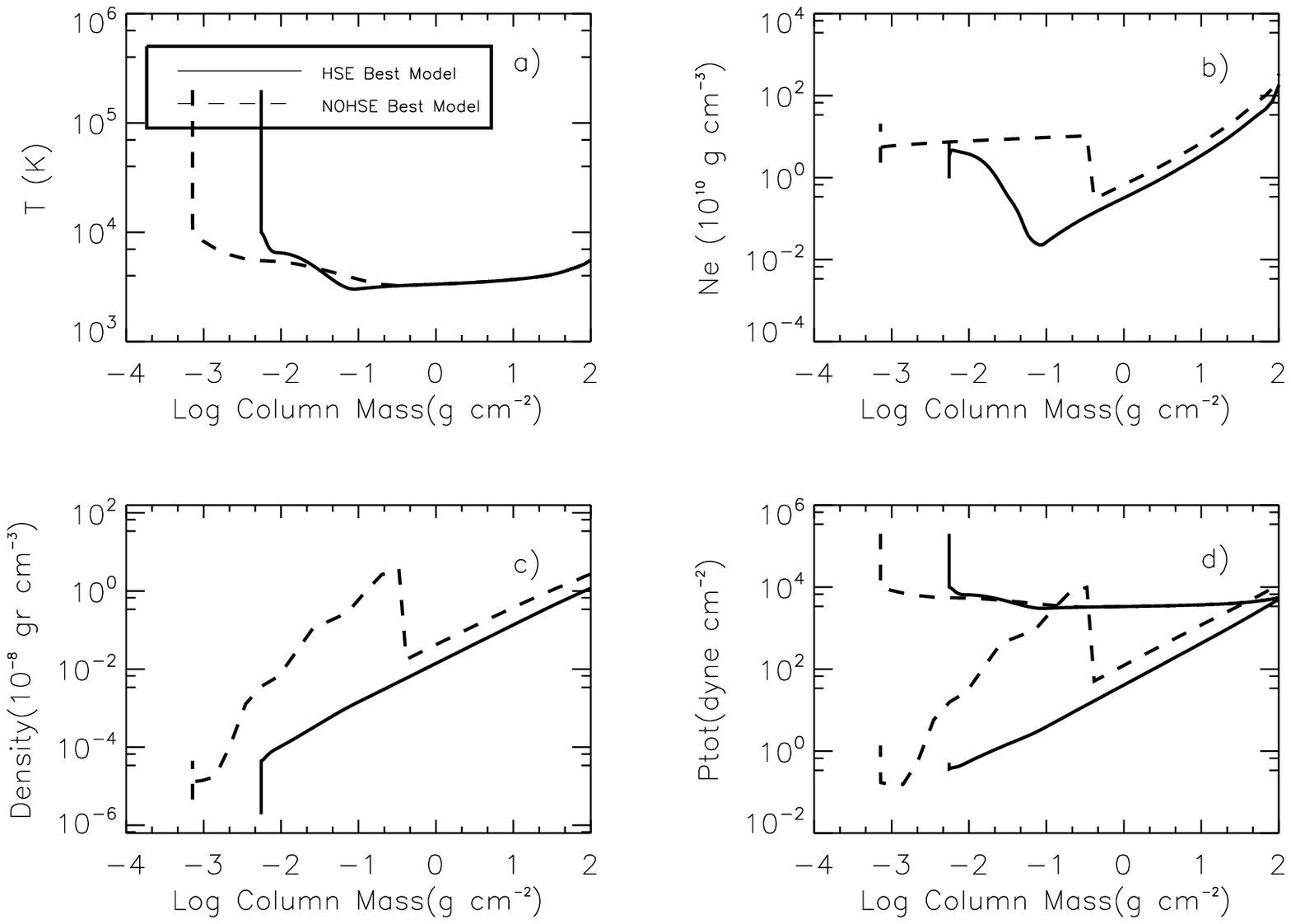}
\caption{Temperature, electron density, density mass and pressure versus Column Mass for the best $HSE$ and $NOHSE$ models of the K2 primary component of the \rscvn\ system HR~7428. In the last plot, the two best models are shown for references.}
\label{best_comparison}
\end{figure*}
This approach takes into account the possibility that some pressure contributions are neglected and becomes a method to derive an estimate of the lacking pressure component. In such an approach we use the electron density as a free parameter, looking for the electron density distribution that best fit the data. We accept as best solution, also a distribution whose total pressure is not balancing the gravity, accepting the hypothesis that a new pressure component could be considered for achieving the equilibrium.

Furthermore, also different approaches in the treatment of line blanketing, produces different $N_e$ vs temperature. It is well known in fact, that the source function of a line can be strongly coupled to radiation fields even at very different wavelengths, via radiative rates determining the statistical equilibrium of the species producing the line. 

 If we would find as the best model reproducing our data, an $HSE$ model, this would mean that magnetic or other contributions to the total pressure are negligible in the whole upper atmosphere of the K2 HR\,7428 star.

In the case where a $NOHSE$ model should best reproduce the observations this would implies that other pressure contributions are important in the outer atmosphere of our star and we can then infer  information on these additive contributions to the total pressure.

Therefore both the $HSE$ grid of 2052 models and the initial grid of 15691 model have been considered in the radiative transfer calculations for \ion{H}{}, \ion{Na}{}, \ion{Ca}{} and \ion{Mg}{}. For each grid, the models that converge to solution for all the atoms have been taken into account for the comparison with the observed spectrum.

The  computed profiles of \halpha, \hbeta, \cairt, \nad, have been convolved with a rotational profile with \vsini=17 \kms \citep{Marino_etal01} and an instrumental profile with R=$\frac {\lambda}{\Delta(\lambda)}$~=~$45,000$, normalized and then compared with observations.

The  computed profiles of \mghk\ have been obtained by the weighted sum of the K2 star profile and the A2 star profile and weighted for the $d^2/R^2$ factor.
Wavelength shifts to account for orbital velocities, are applied to synthetic spectra.

\section{Comparison with observations: the best model}

We used a  $\chi^2$ minimization procedure for the selection of the model that best describes the mean outer atmosphere of HR\,7428.
\begin{figure*}
\includegraphics*[width=0.9\textwidth]{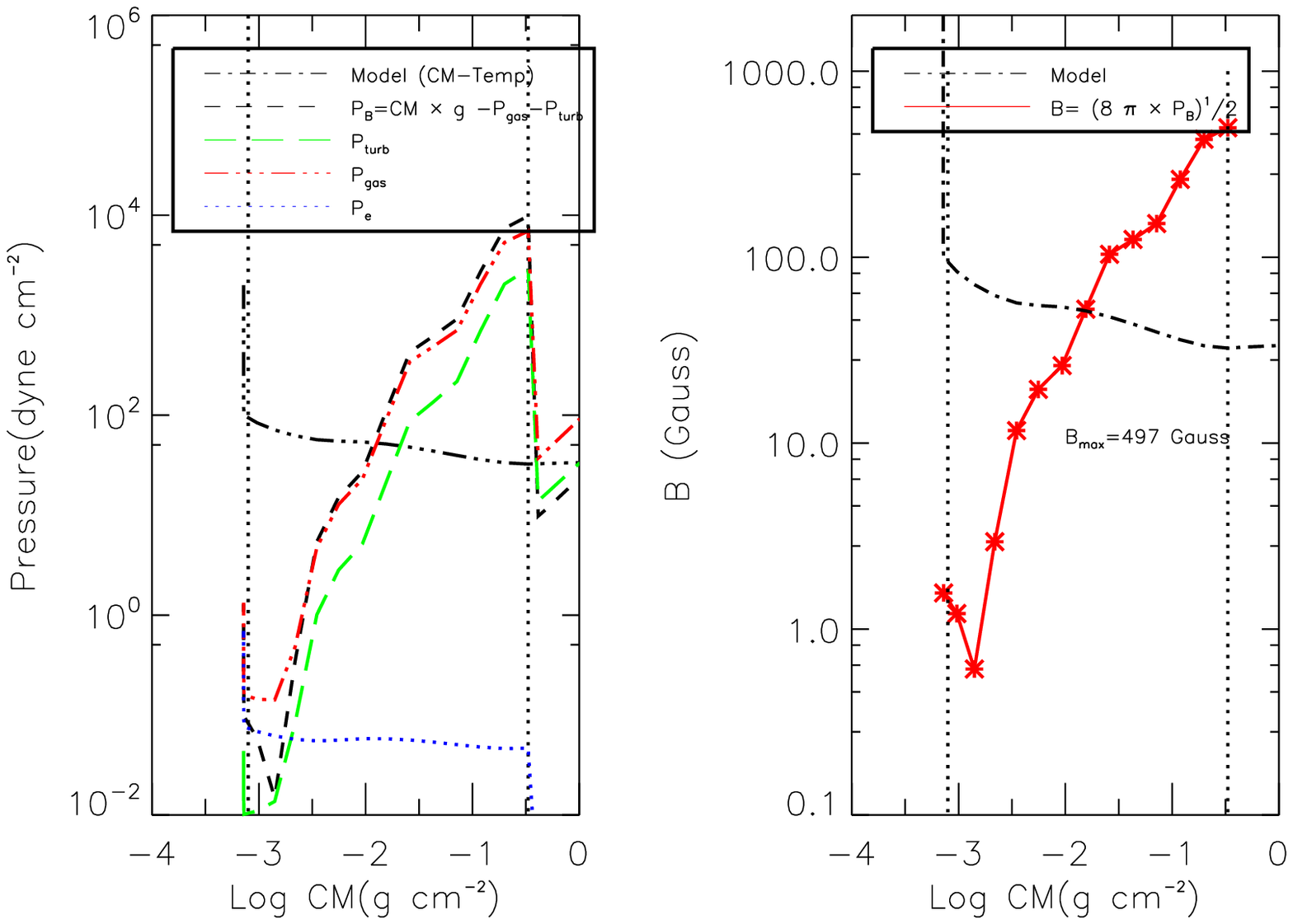}
\caption{(Left panel:)The new pressure component vs Column Mass is shown (dashed line) in comparison to electron pressure (blue), gas pressure (red), turbulent pressure (green); the atmospheric model is overlapped (dot-dashed line) in order to identify the atmospheric layers where we are comparing the pressure components. (Right panel:) Magnetic field distribution vs Column Mass obtained considering the whole lacking pressure as magnetic pressure.}
\label{magneticfield}
\end{figure*}
For each line the $\chi^2$  between each observed and the synthetic line profile has been performed interpolating to the same wavelength grid the two profiles and choosing opportunely the wavelength range for the $\chi^2$ determination. Therefore, for each model, the whole set of observed lines has been compared with the corresponding synthetic lines and a global $\chi^2_{Tot}$ for each model has been calculated as the average of the $\chi^2$ obtained for the single line profiles, a 0.5 weight has been applied to the \mghk\ lines $\chi^2$ in order to take into account the activity-variability as seen in Section~\ref{data}. We defined a selection box formed of those models which give a  $\chi^2$ less than a fixed value for all used lines and UV continuum, the best model is than selected as the one of the box with the lowest $\chi^2_{Tot}$.


We find that the best $NOHSE$ model has a $\chi^2_{Tot}$~=~1.22 while the best $HSE$ model has a $\chi^2_{Tot}$~=~2.60. This result lets conclude that the $NOHSE$ best model distribution of temperature, gas pressure, electron and population densities versus height is the best description of the mean  outer atmosphere of the K2 star of the binary system HR7428. 


We will describe here both the two $HSE$ and $NOHSE$ best models as possible representations of the mean outer atmosphere of the K2 star. 


The best $\chi^2_{Tot}$ of the $NOHSE$ model with respect to the $HSE$ one is clearly understable from Fig.~\ref{righe} where the \halpha, \hbeta, \cairt, \nad\ synthetic profiles computed for the  best $HSE$ (dotted line) and the best $NOHSE$ (dashed line) models are shown and compared with the observed profiles.

The $NOHSE$ model gives \nad\ lines somewhat less deep than the observed profile and an \halpha\ profile a bit deeper than the observed one. Nevertheless, the $NOHSE$ model reproduce much better than the $HSE$  both the \cairt\ line profiles as well as the \hbeta\ profile, furthermore,  the also the $HSE$ model gives \nad\ lines less deep and \halpha\ deeper than the observed one. 

The $NOHSE$ model gives also a better fit of the \mghk\ lines as can be seen from Fig.~\ref{mg} where the IUE observed flux at Earth of the \mghk\ is compared with the synthetic profiles obtained from the $HSE$ (dotted line) and the $NOHSE$ (dashed line) best models. We can see how the $NOHSE$ best model better reproduce both the continuum emission and the \mghk\ line profiles. 
The discrepancy in the peak region of the h and k profiles is compatible with the activity-variability or also to the unresolved ISM absorption line that has not been taken into account in the synthetic lines computation.

Fig.~\ref{cntrbfnohse} and Fig.~\ref{cntrbfhse} show the line contribution functions ($CF$) to the emergent radiation ($CF$) as defined by \cite{Achmad_etal91}, of the synthesized lines in the case of $NOHSE$ and $HSE$ respectively. The plots mainly indicates the region of formation of the different part of each line, having on the $x$ axis the $\Delta \lambda$ from the center of the line and on the $y$ axis the Column Mass that corresponds to an the atmospheric layer. In order to make this more clear we overplotted the atmospheric model as a continuous line in each $CF$ plot.
In the case of \nad, \mghk\ and \cairt, the $CFs$ of the multiple lines are overplotted together. 

We can see that both in the $NOHSE$ and in the $HSE$ atmospheres, the line formation is very similar. 

We find that the base of the transition region is quite well constrained by the  \halpha\ and \mghk\ cores. Also the central part of the \nad, \hbeta\ and \cairt\ cores are affected by the transition region conditions.
\mghk\ lines, except for the core, form entirely in the chromosphere which is also constrained by the \cairt\ lines.
The line wings of all the used lines, except for \mghk\ wings, form in the atmospheric region that goes from the photosphere up to the temperature minimum region.
The last plot on the right, in Fig.~\ref{cntrbfnohse} and Fig.~\ref{cntrbfhse} shows the overplot of all the lines $CF$ and puts in evidence how the used diagnostics enough constrain the whole atmosphere, giving strength to the method and to the atmospheric model derived for the K2 star of HR\,7428.

The $HSE$ and $NOHSE$ best models are shown in Fig.~\ref{bestmodel} compared with the Solar Val~3c model. We can see that the shape of both the $HSE$ and $NOHSE$ best models is quite similar to the Solar one. It is worthwhile to notice that the two models describe strongly different geometries. The $HSE$ best model describes an atmosphere of the mean K2 primary component of the \rscvn\ system HR~7428 that extends up to $8 \times 10^5$~km, the temperature decreases and reaches the minimum temperature of about 3,000~K at about  500,000~Km above the photosphere, the chromosphere extends for 250,000~Km from the temperature minimum up to the base of the transition region that is located at about 800,000~Km. The best $NOHSE$ models is instead less extended of the $HSE$ one; the temperature from the photosphere decreases, reaching the minimum of $\approx$~3200~K at about  140,000~Km above the photosphere, the chromosphere extends for 60,000~Km from the temperature minimum up to the base of transition region that is located at about 200,000~Km. 






The difference between the $HSE$ and the $NOHSE$ models can be better understood from the plot of Fig.~\ref{best_comparison} where the temperature, electron density and total density are plotted versus Column Mass density (CM)  defined as $P/g$ for the best $HSE$ and $NOHSE$ models. 

From Fig.~\ref{best_comparison} panel a), we can see that the $HSE$  best model has a minimum of temperature close to 3000~K when the $\log CM$~$\approx$~-1.04 and a chromospheric 'plateau' at a temperature of about 6500~K at Column Mass $\approx$~$\log CM$~=~-2  then temperature rise up to 10,000~K and at $\log CM$~=~-2.25 we find the base of transition region. In the transition region the temperature rises abruptly up to 200,000~K while the Column Mass remains approximately constant.
The $NOHSE$ best model has a minimum of temperature $\approx$~3250~K (250~K higher with respect to the $HSE$ model) at a $\log CM$ $\approx$~-0.48 a chromospheric plateau colder than the $HSE$ one at about 5500~K and  $\log CM$~=~-2.1. The base of transition region is found here at  $\log CM$~=~-3.14; then temperature rise abruptly up to 200,000~K while the Column Mass remains approximately constant.
It is worthwhile to notice (panel b) that in the chromospheric region the $HSE$ model presents an electron density rising continuously, while in the $NOHSE$ model we observe an abrupt rise of the electron density in a few kilometers while in the main part of the chromosphere the  electron density is approximately constant. This result is the same as that found from semi-empirical models of other cool stars \citep{Harper92}. This abrupt enhancement of electron density corresponds to a density enhancement and to an enhancement of the total pressure. Both in the $HSE$ and in the  $NOHSE$ best models, in the chromospheric layers the density decrease uniformly up to the transition region. Here both electron density and density decrease abruptly.

The most important difference between the two models is shown in Fig.~\ref{best_comparison} panel d), where the total pressure is plotted versus Column Mass. While, in the $HSE$ model, the total chromospheric pressure is imposed to be equal to the Column Mass multiplied for the gravity, that is  $P_{Tot}$~=~CM~$\times g$, and we obtain a straight line, in the $NOHSE$ model, the total chromospheric pressure exceed the gravity pressure (dashed line is not a straight line but, in the chromospheric layers lays well above of the gravity pressure). 

This exceeding pressure has to be balanced by an equal and opposite pressure otherwise we cannot have a stable star. Therefore, we assumed the difference $P_{Tot}$~-~CM~$\times ~g$ as equal, with opposite sign, to the lacking pressure in our calculations, that is: $P_{Totnew}$~=~$P_{turb}$~+~$P_{gas}$~+~$P_{new}$~=~CM~$\times ~g$ and therefore $P_{new}$~=~CM~$\times ~g$~-~$P_{Tot}$. 

It is worthwhile to notice that not imposing hydrostatic equilibrium only refers to the chromospheric layers, both the used photospheric and transition regions models were calculated imposing hydrostatic equilibrium, therefore this new component pressure estimation refers only to this layer.




In Fig.~\ref{magneticfield} the new pressure component is shown in comparison to electron pressure, gas pressure, turbulent pressure. It is clear that this pressure component is not negligible being of the same order of strength of the gas pressure.  In the hypothesis that the additive pressure, could be a magnetic pressure, We calculated a $\approx$~500~gauss magnetic field  corresponding to a magnetic field that, from the base of the chromosphere decrease toward the outer layers. 


\section{Conclusions}
We present here the semi-empirical modeling for the K2 star of the \rscvn\ binary system HR~7428. The model has been computed to match a wide set of observations from the UV continuum to a set of chromospheric lines. The fine coverage in the parameter space used in the modeling let  us able to find a good agreement between the observed and computed spectral features. Furthermore a good agreement obtained in matching the UV continuum when use is made of the line-blanketing approximation method of \cite{Busa_etal01} reinforce the confidence on the method itself.  

Although we have obtained an acceptable agreement between the calculations and the observations  when the $HSE$ is imposed, as it is usual, we do not find an $HSE$ model that has a good match with all the used diagnostics. This could be due to many reasons, here we have explored the possibility that we are neglecting some components to the total pressure in the hydrostatic equilibrium equation. Therefore we considered also the radiative transfers without imposing $HSE$. The best model in $NOHSE$ gives a much better agreement with observations both in line profiles and in UV continuum.The stability of the best $NOHSE$ model, implies the presence of an additive (toward the center of the star) pressure, that decreases  in strength from the base of the chromosphere toward the outer layers. Interpreting this additive pressure as a magnetic pressure we estimated a magnetic field intensity of about 500~gauss at the base of the chromosphere.
 
\section*{Acknowledgements}
We wish to thank the referee, Prof. Donald G. Luttermoser, for his careful reading of the manuscript and for his useful and kind comments and suggestions that we have greatly appreciated. IRAF is distributed by the NOAO which is operated by AURA under contract with NFS.

{}{}


\begin{thebibliography}{}{}
\bibitem[\protect\citeauthoryear{Ayres et al.}{1997}]{Ayres_etal97} Ayres, T.R., Brown, A., Harper, G.M., Bennett, P.D., Linsky, J.L., Carpenter, K.G., \& Robinson, R.D. 1997, ApJ, 491, 876
\bibitem[\protect\citeauthoryear{Ayres et al.}{2003}]{Ayres_etal03} Ayres, T.R., Brown, A., \& Harper, G.M. 2003, ApJ, 598, 610
\bibitem[\protect\citeauthoryear{Achmad et al.}{1991}]{Achmad_etal91} Achmad L.,De Jager C., Nieuwenhuijzen H., 1991 A\&A 250, 445A
\bibitem[\protect\citeauthoryear{Andretta et al.}{2005}]{Andretta_etal2005} Andretta V., Bus\'a I., Gomez M.T., Terranegra L., 2005 A\&A 430, 669
\bibitem[\protect\citeauthoryear{Basri et al.}{1981}]{Basri_etal81} Basri, G.S., Linsky, J.L., \& Eriksson, K. 1981, ApJ, 251, 162
\bibitem[\protect\citeauthoryear{Byrne et al.}{1998}]{Byrne_etal98} Byrne P. B., Abdul Aziz H., Amado P. J., et al., 1998 A\&AS 127, 505
\bibitem[\protect\citeauthoryear{Bus\'a et al.}{2001}]{busaetal99} Bus\'a I., Pagano I., Rodon\'o M., Neff J. E., Lanzafame A. C., 1999 A\&A 350, 571B
\bibitem[\protect\citeauthoryear{Bus\'a et al.}{2001}]{Busa_etal01} Bus\'a I., Andretta V., Gomez M.T., Terranegra L., 2001 A\&A 373, 993
\bibitem[\protect\citeauthoryear{Brown et al.}{1991}]{Brownetal91} Brown A., Drake S. A., Van Steenberg M. E., Linsky J. L., 1991 ApJ, 373, 614B
\bibitem[\protect\citeauthoryear{Cardelli et al.}{1989}]{Cardelli_etal89} Cardelli J. A., Clayton, G. C., Mathis, J. S.,  1989 ApJ, 345, 245C
\bibitem[\protect\citeauthoryear{Carlsson}{1986}]{Carlsson86} Carlsson M., 1986, Technical report 33, Uppsala Astronomical Observatory
\bibitem[\protect\citeauthoryear{Cram \& Mullan}{1979}]{Cram&Mullan79} Cram L. E. \& Mullan D. J., 1979, ApJ, 234, 579
\bibitem[\protect\citeauthoryear{Mullan \& Cram}{1982}]{Mullan&Cram82} Cram L. E. \& Mullan D. J., 1979, ApJ, 234, 579
\bibitem[\protect\citeauthoryear{Fontenla et al.}{1993}]{Fontenla_etal93} Fontenla J. M., Avrett E. H., Loeser R. 1993, ApJ 406, 319
\bibitem[\protect\citeauthoryear{Fontenla et al.}{1999}]{Fontenla_etal99} Fontenla, J., White, O.R., Fox, P.A., Avrett, E.H., and Kurucz, R.L. 1999, ApJ, 518, 480
\bibitem[\protect\citeauthoryear{Gratton}{1950}]{Gratton50} Gratton L. 1950, ApJ 111, 31
\bibitem[\protect\citeauthoryear{Griffiths \& Jordan}{1998}]{GriffithsJordan98} Griffiths N. W., Jordan C., 1998, ApJ,497, 883
\bibitem[\protect\citeauthoryear{Jordan \& Brown}{1981}]{JordanBrown81} Jordan C. \& Brown A., 1981 SPSS, 199J
\bibitem[\protect\citeauthoryear{Hall et al.}{1990}]{Hall_etal90} Hall D. S., Gessner S. E. Lines H. C., Lines R. D., 1990 AJ 100, 2017
\bibitem[\protect\citeauthoryear{Harper}{1992}]{Harper92} Harper G. M., 1992 MNRAS 256, 37
\bibitem[\protect\citeauthoryear{Houdebine}{1996}]{Houdebine_96} Houdebine E. R., 19960 IAUS 176, 547H
\bibitem[\protect\citeauthoryear{Kalkofen et al.}{1998}]{Kalkofen_etal99} Kalkofen W., Ulmschneider, P., Avrett E. H., 1999 ApJ 521, 141
\bibitem[\protect\citeauthoryear{Kelch et al.}{1978}]{Kelch_etal78} Kelch, W.L., Linsky, J.L., Basri, G.S., Chiu, H-Y, Chang, S-H, Maran, S.P., \& Furenlid, I. 1978, ApJ, 220, 962
\bibitem[\protect\citeauthoryear{Kurucz}{1993}]{Kurucz93} Kurucz R. L. 1993, A new opacity-sampling model atmosphere program for arbitrary abundances, in Peculiar versus normal phenomena in A-type and related stars, IAU Coll. 138, ed. M. M.Dworetsky, F. Castelli, R. Faraggiana, ASP Conf. Ser., 44, 87
\bibitem[\protect\citeauthoryear{Kurucz \& Avrett}{1981}]{Kurucz_Avrett_81} Kurucz R. L., Avrett E.H, 1981, SAOSR, 391, Solar Spectrum Synthesis: A Sample Atlas from 224 to 300 nm 
\bibitem[\protect\citeauthoryear{Lanzafame et al.}{2000}]{Lanzafame_etal00} Lanzafame A. C., Bus\'a I., Rodon\`o M., 2000, A\&A 362, 683
\bibitem[\protect\citeauthoryear{Leone et al.}{2016}]{leone_etal16} Leone F., Avila G., Bellassai G., et al., 2016, AJ, 151, 116L
\bibitem[\protect\citeauthoryear{Linsky et al.}{2001}]{linsky_01} Linsky J., Redfield S., Ayres T., Brown A., Harper G., 2001, ASPC, 242, 247L
\bibitem[\protect\citeauthoryear{Luttermoser et al.}{1994}]{Luttermoser_etal94} Luttermoser, D.G., Johnson, H.R., \& Eaton, J.A. 1994, ApJ, 422, 351
\bibitem[\protect\citeauthoryear{Marino et al.}{2001}]{Marino_etal01} Marino G., Catalano S., Frasca A., Marilli E., 2001, A\&A 375, 100
\bibitem[\protect\citeauthoryear{Mauas \& Falchi}{1994}]{Mauas&Falchi94} Mauas P. J. D. \& Falchi A., 1994, A\&A, 281, 129
\bibitem[\protect\citeauthoryear{Mauas et al.}{1997}]{Mauas_etal97} Mauas P. J. D., Falchi A., Pasquini L., Pallavicini R., 1997, A\&A 326, 249
\bibitem[\protect\citeauthoryear{Mauas et al.}{2006}]{Mauas_etal06} Mauas P. J. D., Cacciari C., Pasquini L., 2006, A\&A 454, 609
\bibitem[\protect\citeauthoryear{Pagano et al.}{2006}]{Pagano_etal06} Pagano I., Ayres T. R., Lanzafame A. C., Linsky J. L., Montesinos B., Rodon\'o M., 2006 Ap\&SS 303,17
\bibitem[\protect\citeauthoryear{Parsons \& Ake}{1987}]{Parsons&Ake87} Parsons S. B., Ake T. B. 1987, Bull. Am. Astron. Soc. 19, 708
\bibitem[\protect\citeauthoryear{Robinson et al.}{1998}]{Robinson_etal98} Robinson, R.D., Carpenter, K.G., \& Brown, A. 1998, ApJ, 503, 396
\bibitem[\protect\citeauthoryear{Short \& Doyle}{1998}]{Short&Doyle98} Short C. I. \& Doyle J. G., 1998, A\&A, 336, 613
\bibitem[\protect\citeauthoryear{Span\'o et al.}{2006}]{Spano_etal06} Span\'o P., Leone F., Bruno P., Catalano S., Martinetti E., Scuderi S., 2006, MSAIS, 9, 481S
\bibitem[\protect\citeauthoryear{Span\'o et al.}{2004}]{Spano_etal04} Span\'o P., Leone F., Scuderi S., Catalano S., Zerbi F. M., 2004, SPIE, 5492, 373S
\bibitem[\protect\citeauthoryear{Uitenbroek}{1998}]{Uitenbroek92} Uitenbroek H., 1992, ASPC, 26, 546 
\bibitem[\protect\citeauthoryear{Vernazza et al.}{1973}]{Vernazza_etal73} Vernazza J. E., Avrett E. H., Loeser R., 1973, ApJ 184,605V
\bibitem[\protect\citeauthoryear{Vernazza et al.}{1981}]{Vernazza_etal81} Vernazza J. E., Avrett E. H., Loeser R., 1981, ApJS 45,635
\bibitem[\protect\citeauthoryear{Vieytes et al.}{2005}]{Vieytes_etal05} Vieytes M.; Mauas P.; Cincunegui C., 2005, A\&A, 441, 701
\end{thebibliography}
\end{document}